\documentclass{IEEEtran}
\IEEEoverridecommandlockouts
\usepackage{cite}
\usepackage{algorithmic}
\usepackage{xcolor}
\usepackage{enumerate}
\usepackage{amsfonts,amsmath,amssymb,amsfonts}
\usepackage{amsthm}
\usepackage{algorithmic}
\usepackage{optidef}
\usepackage{algorithm}
\usepackage{mathrsfs}
\usepackage{graphicx}
\usepackage{textcomp}
\usepackage{subfigure}
\usepackage[squaren]{SIunits}
\usepackage[section]{placeins}

\def\BibTeX{{\rm B\kern-.05em{\sc i\kern-
.025em b}\kern-.08em
    T\kern-.1667em\lower.7ex\hbox{E}\kern-.125emX}}
\begin{document}
\newtheorem{lemma}{\textbf{Lemma}}
\bibliographystyle{IEEEtran}

\title{Resource Optimization for Semantic-Aware
Networks with Task Offloading}
\author{\IEEEauthorblockN{Zelin Ji,~\IEEEmembership{Student Member,~IEEE},
Zhijin Qin,~\IEEEmembership{Senior Member,~IEEE}, Xiaoming Tao,~\IEEEmembership{Senior Member,~IEEE}, and \\ Zhu Han,~\IEEEmembership{Fellow,~IEEE}}
\thanks{Part of this work was presented at the IEEE International Conference on Communications 2023~\cite{ji2023edge}.}
\thanks{Zelin~Ji is with School of Electronic Engineering and Computer Science, Queen Mary University of London, London E1 4NS, U.K. (email: z.ji@qmul.ac.uk).}
\thanks{Zhijin~Qin and Xiaoming~Tao are with Department of Electronic Engineering, Tsinghua University, Beijing, China. (email: qinzhijin@tsinghua.edu.cn; taoxm@tsinghua.edu.cn).
}
\thanks{
Zhu~Han is with the University of Houston, Houston, TX 77204 USA. (e-mail: zhan2@uh.edu).
}}
\maketitle

\begin{abstract}
The limited capabilities of user equipment restrict the local implementation of computation-intensive applications. Edge computing, especially the edge intelligence system, enables local users to offload the computation tasks to the edge servers to reduce the computational energy consumption of user equipment and accelerate fast task execution. However, the limited bandwidth of upstream channels may increase the task transmission latency and affect the computation offloading performance. To overcome the challenge arising from scarce wireless communication resources, we propose a semantic-aware multi-modal task offloading system that facilitates the extraction and offloading of semantic task information to edge servers. To cope with the different tasks with multi-modal data, a unified quality of experience (QoE) criterion is designed. Furthermore, a proximal policy optimization-based multi-agent reinforcement learning algorithm (MAPPO) is proposed to coordinate the resource management for wireless communications and computation in a distributed and low computational complexity manner. Simulation results verify that the proposed MAPPO algorithm outperforms other reinforcement learning algorithms and fixed schemes in terms of task execution speed and the overall system QoE.

\end{abstract}


\section{Introduction}

The significant surge in wireless data traffic has led to an escalation in computational demands on both core networks and local devices, exerting immense pressure on wireless communication systems. In response to these challenges, novel techniques such as edge computing \cite{9606720} have emerged to offload the computation burden from the core network and local devices to the edge servers (ESs). Furthermore, semantic communications \cite{qin2022semantic} have been proposed and developed, which are robust to fast varying channel environments and can reduce the size of transmitted data. By extracting the meanings of data and filtering out useless and irrelevant information, semantic communications are expected to transmit only essential task-related information, which makes it a promising technique for task offloading systems.

\subsection{Resource Allocation in Edge Intelligence}

The limited resources in wireless communications have become a bottleneck in data transmission between local devices and ESs, significantly impacting the performance of edge computing. To minimize the communication latency and improve computational capability, various studies have focused on resource allocation techniques. These techniques optimize the transmit power~\cite{9210812}, user scheduling~\cite{8851249}, and bandwidth allocation~\cite{9194337} for edge intelligence. Zhou \emph{et al.}~\cite{9556549} consider both communication and computation resources for local devices and ESs, jointly optimizing computational capability and transmit power of local users to achieve energy-efficient resource allocation. However, these works rely on stable channel state information (CSI), which is impractical for 5G, 6G and beyond communication systems. {To overcome the challenge of the fast-varying channels and make decisions under unstable CSI, Min~\emph{et al.}~\cite{8598893} proposed a reinforcement learning (RL) based offloading scheme for an IoT device without knowledge of the servers.} Meanwhile, the presence of fast-varying channels and limited bandwidth resources necessitates communication techniques that are robust to channel variations while reducing data traffic for task offloading.

\subsection{Semantic-aware Task Offloading Systems}

Semantic communications have recently emerged as a promising solution to address the challenges mentioned above. Yang \emph{et al.}~\cite{yang2022semantic} have explored semantic extraction procedures to enhance edge intelligence. In contrast to traditional communications, semantic communications transmit and receive the meaning of raw data instead of converting it into bit streams. This approach is robust to channel variations and requires a lower signal-noise-ratio (SNR) for accurate data recovery~\cite{qin2022semantic}. Specifically, Xie \emph{et al.}~\cite{9830752} have demonstrated that certain specific tasks can be executed based on the retrieved semantic information, inspiring us to compress local tasks into semantic information and transmit the compressed semantic data to the ESs, thus reducing the data traffic and enhancing the robustness of the communication system significantly.

Some studies have made progress in applying deep learning based semantic communications on text~\cite{9252948,10000901}, image~\cite{image}, audio~\cite{speech}, video~\cite{video}, and multi-modal data~\cite{qin2022semantic,unified}. Different types of tasks are based on multi-modal data, which require different semantic transceiver designs. Moreover, different user equipment (UEs) have preferences on various metrics, e.g., task execution accuracy, task execution latency, or energy consumption, which are the key requirements to realize edge intelligence in 6G communications. Conventional resource allocation approaches usually focus on a single type of task and optimize a global goal, which neglects the personality among UEs.

Furthermore, existing semantic applications rely on large deep learning models that are computationally heavy, making it challenging to implement them on the end device side. For example, running a traditional model~\cite{LXMERT} for the visual question answering (VQA) task on the mobile board takes nearly six seconds. To provide seamless usage without perceivable delays, task latency should be processed and executed within one second. To address this latency issue, Cao~\emph{et al.}~\cite{mobivqa} proposed a simplified VQA model that can be implemented on the mobile phone, where the local models process the VQA tasks fast and with some accuracy loss. Alternatively, offloading tasks to edge servers offers a promising solution. By converting high computational costs to compressed semantic communication overhead, the computation burden at local UEs can be offloaded to ESs with high computational capacities. This reduces local energy consumption and task execution latency.

User privacy leakage is a significant concern in edge computing. In traditional edge computing systems, an eavesdropper can intercept the source message transmitted by the transmitter, compromising user privacy. {He~\emph{et al.}~\cite{8491311} designed the privacy metric to ensure that the devices can maintain the level of privacy during offloading. Min~\emph{et al.}~\cite{8491311} applied an RL based privacy-aware offloading scheme to protect healthcare privacy.} On the other hand, Du~\emph{et al.}~\cite{du2022rethinking} have conducted an analysis of potential scenarios where user privacy can be protected using semantic communications. These semantic communication models are trained end-to-end with cross-layer co-design. Even if the eavesdropper can receive the compressed semantic features of UEs, it is difficult to decode the source information due to the lack of background knowledge and cross-layer training. The decoding of semantic information depends on the characteristic and private design of the receivers, making semantic communications a potential approach for achieving secure communications.

In this paper, a semantic-aware multi-task offloading system is presented. The high local computation loads of the tasks can be converted to semantic information with a low communication load, which can then be offloaded to the ESs with a sufficient power supply. Moreover, we employ the quality of experience (QoE) as the optimization criterion, meanwhile considering the preferences of UEs. This QoE-based resource management is adopted to perform a joint resource allocation for the communication cost and the computation cost, thus maximizing the QoE preference among UEs. To demonstrate the effectiveness of the proposed resource management for the semantic-aware multi-task offloading system, an edge intelligence system with multi-users and heterogeneous task sets needs to be considered. To the best of our knowledge, this is the first paper that performs the joint optimization of the computation resources and communication resources for semantic-aware networks with multi-modal tasks. Nevertheless, to realize the aforementioned techniques, there are some challenges that we need to tackle.

\begin{itemize}
 \item {How to implement a semantic-aware multi-task offloading system?}

\item {How to fairly measure the computational cost for different semantic tasks?}

\item {Why do we need to investigate the semantic-aware multi-task offloading system? Is it necessary to offload the tasks to the edge server?}

\item {How to jointly optimize the computational costs and communication costs to maximize the preferred task execution QoE of UEs?}
\end{itemize}

To address the above challenges, the contributions of this paper are summarized as follows.
\begin{enumerate}
\item {A semantic-aware multi-task offloading system is considered. Instead of offloading the tasks to the ESs directly, we employ deep learning enabled semantic communication (DeepSC) systems to extract the multi-modal task semantics and transmit the compressed, low-dimensional information to the ESs. }

\item {We consider machine translation, image retrieval, and VQA tasks as examples to implement the semantic-aware multi-task offloading system. The text and image semantic encoders are based on the popular Transformer models~\cite{vaswani2017attention, touvron2021training}. The quantified computational costs analysis of executing the above models is provided to verify the necessity of the task offloading.}

\item {To evaluate the performance of the semantic task offloading system for multi-modal data, the paper introduces QoE metrics for task execution based on the task execution cost and the execution accuracy. Moreover, to fairly evaluate the execution cost for multi-modal tasks, the execution cost of the time latency and energy consumption are standardized by the local execution scheme.}

\item {The joint management for computation resources and communication resources is performed distributedly using the proximal policy optimization based multi-agent reinforcement learning (MAPPO) algorithm at the local UE. The local UE can make decisions based on the local information, i.e., GPS location, battery life, and task queue, hence relieving the computational pressure of the cloud base station (BS) and reducing the local information uploading overhead.}
\end{enumerate}

The rest of this article is organized as follows. Section~\ref{sec:system_model} presents the transmission and computation models for the multi-modal task offloading system. In Section~\ref{sec:problem}, the criteria of the multi-modal tasks are designed and the QoE maximization problem is formulated. The MAPPO algorithm to solve the formulated problem is presented in Section~\ref{sec:algorithm}, where the detailed training process is introduced. The numerical results are illustrated in Section~\ref{sec:numerical_results}. Finally, the conclusion is drawn in Section~\ref{sec:conclusion}.

\section{System Model}
\label{sec:system_model}
\subsection{Network Model}
\begin{figure*}[t]
\centering
\includegraphics[width=2\columnwidth]{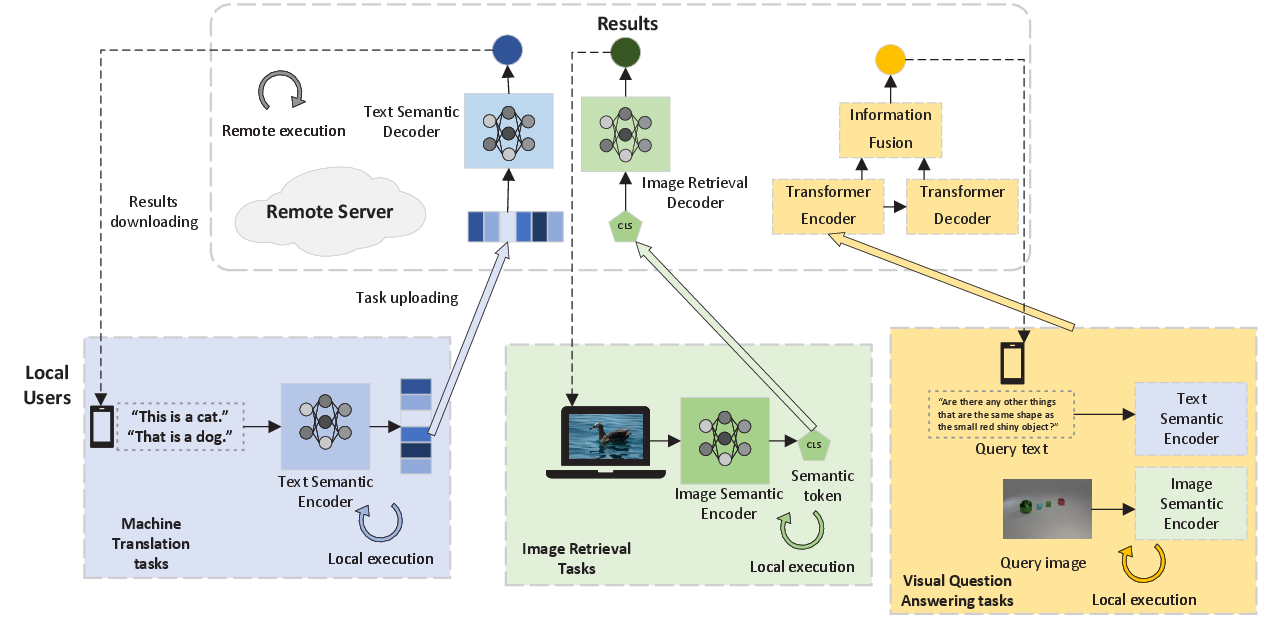}
\caption{System model for semantic task offloading process.}
\label{system_model}
\end{figure*}

To reduce the transmission cost of the offloading tasks from local users to the edge server, we adopt the DeepSC model proposed by Xie~\emph{et al.}~\cite{9252948}. The semantic communications are implemented based on the pre-trained semantic encoder and decoder. We define the set of UEs as ${\mathcal U} = \{U_i|i=1,\dots,I\}$. The $i$-th UE can associate with the ES to offload the machine translation tasks, i.e., the associate coefficient $\rho_i = 1$, or perform the task locally, where $\rho_i = 0$.

Generally, the semantic task offloading processes include several steps. Firstly, $U_i$ offloads the semantic task with modality $\mathcal Q = \{{\mathcal I}: \rm{image}, {\mathcal T}: \rm{text}, {\mathcal B}: \rm{bimodal}\}$ as $T^{\mathcal Q}_i = \{d^{\mathcal Q}_i, l^{\mathcal Q}_i, \tau_{\rm{max}\}}$ to the ES, where $d^{\mathcal Q}_i$ denotes the number of tasks in the queue, $l^{\mathcal Q}_i$ is the average hardware requirement per task\footnote{Note that we consider the floating-point operations per second (FLOPS) computing capability for BERT models, which are usually performed by the graphics processing unit (GPU).}, and $\tau_{\rm{max}}$ is the maximize latency constraint for a task. Then the semantic tasks and the computation are performed by the ESs. In the third phase, the results for the semantic tasks are downloaded from the ESs to the UEs. The process is illustrated in Fig.~\ref{system_model}. Note that the joint semantic and channel (JSC) encoders and JSC decoders are also implemented to map the extracted semantic features to the complex signals which can be transmitted by the physical channels. The JSC encoders and decoders are also pre-trained by the end-to-end networks and implemented for the task offloading system, while we omit these blocks to simplify the Fig.~\ref{system_model}.

\subsection{Transmission Model}
Consider an orthogonal frequency-division multiple access (OFDMA) based semantic-aware network with an ES. The ES can communicate with $U_i$ on the bandwidth $b_i$. Suppose that all channels consist of large-scale fading and small-scale fading, and the signal-to-noise ratio (SNR) of $U_i$ can be denoted as 


\begin{equation}
    \gamma_{i} = \frac{\rho_{i}{p_i}{g_i}|{h_i}|^2}{b_i \sigma^2},
\end{equation}
where $p_i$ is the transmit power, $g_i$ represents the large-scale channel gain and $h_i \thicksim \mathcal{CN}(0,1)$ is the small-scale fading coefficient for the sub-channel assigned to $U_i$, $\sigma^2$ is the noise power spectral density.

\subsection{Task Offloading Model}
Unlike the bit-stream data rate for the traditional Shannon paradigm, the semantic rate is based on the semantic unit and the amount of semantic information. The input source data of the semantic encoder is denoted as $\boldsymbol{x^{\mathcal Q}_i}=\{x[0], x[1], \dots, x[S^{\mathcal Q}_i]\}$, where $S^{\mathcal Q}_i$ represents the length of the input data. For example, for a text semantic encoder, the input $\boldsymbol{x^{\mathcal T}_i}$ is a sentence, $x[\cdot]$ represents the words in the input sentence, while $S^{\mathcal T}_i$ is the number of words in a sentence. We define the average semantic information for the input $\boldsymbol{x^{\mathcal Q}_i}$ as $A^{\mathcal Q}_i$ according to~\cite{9763856}, and the semantic rate can be expressed by 
\begin{equation}
    \Gamma_i = \frac {b_i A^{\mathcal Q}_i \epsilon^{\mathcal Q}_i}{S^{\mathcal Q}_ik^{\mathcal Q}_i},
\end{equation}
where $k^{\mathcal Q}_i$ denotes the average number of semantic symbols used for each input unit $x[\cdot]$, and $\epsilon^{\mathcal Q}_i$ is the task execution accuracy that is related to $k^{\mathcal Q}_i$ and SNR $\gamma_i$, which will be discussed later. Note that $A^{\mathcal Q}_i$, $S^{\mathcal Q}_i$, and $k^{\mathcal Q}_i$ are the parameters that depend on the pre-trained DeepSC model and the source type, and can be considered constant values during the resource optimization process.

\subsection{Computational Model}
For the tasks performed locally, i.e., $\rho_i = 0$, the computational latency can be expressed by

\begin{equation}
    t^{lc}_i = (1-\rho_i)\frac{d^{\mathcal Q}_i l^{\mathcal Q}_i}{c_i},
\end{equation}
where $d^{\mathcal Q}_i$ and $l^{\mathcal Q}_i$ are task-related parameters as defined before, $c_i = n_if_i$ is the theoretical GPU computation capability, $f_i$ is the GPU clock frequency, and $n_i$ is the number of floating-point operations that the local GPU can execute per cycle of $U_i$. The energy consumption for performing semantic tasks locally can be represented as $E^{lc}_i = \alpha^{lc} t^{lc}_i (f_i)^3$.

For tasks offloaded to ESs, i.e., $\rho_i = 1$, the $U_i$ is associated with the ES, and the machine translation tasks are offloaded to the ES for remote computation. The transmit latency is given by:

\begin{equation}
    t^{ut}_i={\rho_id^{\mathcal Q}_iA^{\mathcal Q}_i}/{\Gamma_i}={\rho_id^{\mathcal Q}_iS^{\mathcal Q}_ik^{\mathcal Q}_i}/{b_i\epsilon^{\mathcal Q}_i},
\end{equation}
where the energy consumption of communication can be expressed as $E^{ut}_i=p_it^{ut}_i$. For the machine translation tasks performed remotely, the time latency can be further expressed as:
\begin{equation}
t^{rc}_i = \frac{d^{\mathcal Q}_i l^{\mathcal Q}_i}{c^{rc}/\sum \nolimits_{{i \in I}} \rho_i},
\end{equation}
where $c^{rc} = n^{rc}f^{rc}$ is the GPU computation capability frequency of the ES, $n^{rc}$ is the number of floating-point operations that the remote GPU can execute per cycle, and $f^{rc}$ is the clock frequency of the remote GPU. Thereby, the energy consumption by $ES$ can be expressed as
\begin{equation}
E^{rc}_i = \alpha^{rc} t^{rc}_i ({f^{rc}}/{\sum \nolimits_{{i \in I}}\rho_i})^3,
\end{equation}
where $\alpha^{rc}$ is the remote energy related coefficient.

The overall task perform latency for $U_i$ can be therefore expressed as
\begin{equation}
    t_i = \rho_i (t^{ut}_i+t^{rc}_i+t^{dl}_i)+(1-\rho_i)t^{lc}_i,
\end{equation}
where $t^{dl}_i$ denotes the results downloading latency. Note that the upload bandwidths for the semantic tasks are relatively small and the downstream bandwidth is higher, and thus the downloading latency can be treated as a small constant. Considering that there is sufficient power supply at the ES side, we only consider the energy consumption on the UE side
\begin{equation}
    E_i=E^{lc}_i+E^{ut}_i + E^c_i,
\label{energy_consumption}
\end{equation}
where $E^c_i = t_i p^c_i$ represents the static and circuit energy for maintaining the basic operations of the UE system. In the previous works, Ji~\emph{et al.}~\cite{ji2023edge} minimized the sum energy consumption in (\ref{energy_consumption}) of the UEs for the single modal machine translation task. However, it is unfair to compare the task execution latency and energy consumption directly for multi-modal tasks with various resource requirements. The following section analyzes the metrics in detail and a unified and fair QoE criterion is designed.

\section{Multi-modal Task Offloading Design and Problem Formulation}
\label{sec:problem}
The investigation of tasks involving single-modal and bimodal data is crucial for the semantic-aware multi-modal task offloading paradigm. Based on the analysis presented above, it is necessary to define a unified criterion that takes into account the QoE of the UEs. Particularly, the diverse preferences of UEs in terms of energy consumption, task execution latency, and task execution accuracy should be considered. Using the analysis and definitions provided, we formulate the problem of maximizing QoE as a joint optimization of the offloading policy, the offloading transmit power, and the local GPU frequency.


We begin by discussing the process of text-based task offloading. In this case, the machine translation task is performed using the bidirectional encoder representations from the Transformers model~\cite{vaswani2017attention}, and we quantify the computational energy consumption of this model. Let $U_i$ have a machine translation task $T^{\mathcal T}_i$, meaning $d^{\mathcal T}_i$ sentences need to be translated, and $l^{\mathcal T}_i$ represents the hardware requirement for translating a sentence. The sentences to be translated are input as $\boldsymbol{x^{\mathcal T}_i}$ into the DeepSC model to extract the semantic information, resulting in an average of $S^{\mathcal T}_i$ words and $A^{\mathcal T}_i$ semantic information. The machine translation accuracy can be measured by the Bilingual Evaluation Understudy (BLEU) score. The task offloaded execution accuracy is defined as the ratio of the offloaded BLEU score to the local BLEU score. i.e., the ratio of the BLEU score of the task executed by the remote server, to the BLEU score executed locally for the same sentence\footnote{It is important to note that we use the standardized accuracy instead of the BLEU score to ensure fair measurement of the quality of experience for different modalities of data.}. The accuracy $\epsilon^{\mathcal T}_i$ can be directly obtained from the pre-trained models and is related to $k^{\mathcal T}_i$, the number of symbols per word, and the signal-to-noise ratio (SNR) $\gamma_i$, i.e., $\epsilon^{\mathcal T}_i = f^{\mathcal T}(k^{\mathcal T}_i, \gamma_i)$.

For image-based tasks, we consider the case of image retrieval tasks. Let $d^{\mathcal I}_i$ denote the number of images in the task queue, and similarly, let $l^{\mathcal I}_i$ represent the hardware requirement for classifying an image. An image is decomposed into $S^{\mathcal I}_i$ patches, with $A^{\mathcal I}_i$ semantic information, including an image semantic token. The accuracy for the image retrieval task is defined as the ratio of the offloaded classification accuracy to the local accuracy. Similar to the text-based tasks, the accuracy $\epsilon^{\mathcal I}_i$ can be obtained from the pre-trained models and is matched with the output dimension and the SNR, i.e., $\epsilon^{\mathcal I}_i = f^{\mathcal I}(k^{\mathcal I}_i, \gamma_i)$.

In the case of VQA tasks, which involve both text and image modalities, a query sentence and a query image need to be processed. The local UEs preprocess the query text using a text semantic encoder and the query image using an image semantic encoder. The VQA tasks can be executed either by the local VQA model or the cloud VQA model~\cite{mobivqa}, and the offloading policy can be determined by the local UEs. The time latency and energy consumption of the VQA tasks can be calculated as the sum of the text and image tasks described above. The task execution accuracy for VQA tasks is related to the text semantic encoder, the image semantic encoder, and the SNR of the local UEs, i.e., $\epsilon^{\mathcal B}_i = f^{\mathcal B}(k^{\mathcal T}_i, k^{\mathcal I}_i, \gamma_i)$.

The objective of this paper is to maximize the task execution performance of the semantic offloading system. Without loss of generality, the task execution cost is assumed to have a linear dependence on both the energy consumption and the task execution latency. However, it is unreasonable to directly compare the energy consumption and the task execution latency for multi-modal data. Moreover, simply minimizing either the energy consumption or the time latency for the entire system may have adverse effects on fairness. To address these issues, we define the standardized task execution cost for $U_i$ as
\begin{equation}
    \rm{Cost}_i = {\beta}\frac{E_i}{E^{\mathcal Q}_0}+(1 - \beta)\frac{t_i}{t^{\mathcal Q}_0},
\label{cost}
\end{equation}
where $\beta$ is the QoE preference coefficient for the energy consumption. Additionally, $E^{\mathcal Q}_0$ and $t^{\mathcal Q}_0$ represent the energy consumption and latency for the task executed locally, thus we can measure the QoE for different UEs fairly. Generally speaking, the QoE is negatively correlated with the cost while proportional to the task accuracy, which is denoted as {
\begin{equation}
    \rm{QoE}_i = \frac{\epsilon^{\mathcal Q}_i}{\epsilon^{\mathcal Q}_0\cdot\rm{Cost}_i},
\label{qoe}
\end{equation}}
where $\epsilon^{\mathcal Q}_0$ is the local execution accuracy for $\forall {\mathcal T}$, $\forall {\mathcal I}$, and $\forall {\mathcal B} \in {\mathcal Q}$. Here we standardize the task QoE with the tasks executed locally. 

To maximize the overall QoE of the user group, we consider the joint optimization of the computation resources and the communication resources. Particularly, the user can determine the offloading policy, the local processing frequency, the offloading power, and the bandwidth requirement. Mathematically, the optimization problem can be formulated as
\begin{maxi!}|l|
{\{\boldsymbol \rho, \boldsymbol p, \boldsymbol f, \boldsymbol b\}}{\sum^I_{i=1}\rm{QoE}_i}
{\label{eq20}}{(\textbf{P0})}
\addConstraint{\epsilon^{\mathcal Q}_i \geq \epsilon_{min} \label{objective:c1} }
\addConstraint{\rho_i = \{0,1\} \label{objective:c2} }
\addConstraint{p_i \leq p_{\rm{max}} \label{objective:c3} 
}
\addConstraint{f_i \leq f_{\rm{max}} \label{objective:c4} 
}
\addConstraint{b_i \leq BW \label{objective:c5} 
}
\addConstraint{t_i \leq t_{\rm{max}} \label{objective:c6} 
}
\addConstraint{E_i \leq E^{\rm{max}}_i \label{objective:c7} 
}
\addConstraint{i = 1,2,\dots, I, \label{objective:c8} 
}
\end{maxi!}
where $\{\boldsymbol \rho, \boldsymbol p, \boldsymbol f, \boldsymbol b\} = \{\rho_i, p_i,f_i, b_i\}, \forall i \in \{1,\dots, I\}$, $\epsilon_{min}$ represents the minimum semantic accuracy requirement, $p{_{\rm{max}}}$ is the maximum transmit power constraint, $f_{\rm{max}}$ is the maximum boost GPU clock frequency, $BW$ is the overall bandwidth for the user group $\mathcal I$, $t_{\rm{max}}$ is the maximum latency constraint, $E^{\rm{max}}_i$ is the maximum battery capacity of $U_i$, which guarantee the battery life during the task processing time. If all constraints are satisfied, the task can be executed successfully, and the task queue $d^{\mathcal Q}_i \rightarrow d^{\mathcal Q}_i -1$. 

Traditionally, the computation offloading strategy can be determined by Lyapunov optimization by decoupling the joint optimization problem to sequential per-stage deterministic subproblems~\cite{8638800}. However, these algorithms are usually executed at the central BS or the ESs, which increases the communication cost and computation pressure of the center. Moreover, it is unreasonable for the center to control the transmit power and GPU clock frequency of the massive UEs. Therefore, to perform online optimization distributedly and reduce the complexity of the joint optimization algorithm, we proposed the MAPPO algorithm, introduced in the following section.

\section{Multi-agent Proximal
Policy Optimization for Resource Allocation Algorithm}
\label{sec:algorithm}
In this section, we present the components of the MAPPO algorithm, starting with the definition of the state, observation state, action space, and reward. Subsequently, we propose an advanced network structure and describe the training process for the MAPPO. Each UE is treated as an RL agent, capable of making decisions independently based on the local RL model.

\subsection{MAPPO Components Definition}

To make the task offloading decision, while achieving the objective and satisfying the constraints, UE needs to consider its battery life, the channel state information, and the computation requirement of the tasks. However, as a local agent, UE cannot observe the global state. Mathematically, the local observation space for the $U_i$ can be expressed as $o_i = \{\boldsymbol{s}_i, E^{\rm{max}}_i, t_i\}$, where $\boldsymbol{s}_i$ is the location of the $U_i$. {According to the local observation, the local UEs need to determine if the task needs to be offloaded in each time step. }

The UEs are permitted to access subchannels to offload the tasks to the ES. However, the OFDM scheme cannot deal with the situation of two UEs accessing the same subchannel. To avoid the collision issue, we assume that each UE can only determine its bandwidth requirement $\hat{b}_i$, i.e., it cannot determine which subchannel to access. The ES collects the bandwidth requirement information and allocates the bandwidth according to 
\begin{equation}
    b_i = BW \frac{\mathrm{e}^{\hat{b}_i}}{\sum^I_{i=1}\mathrm{e}^{\hat{b}_i}}.
\end{equation}

For the tasks executed locally, the transmit power and the bandwidth requirement are set to a negligible value. The action space includes the variables that we want to optimize, i.e., the offloading policy, the bandwidth requirement, the offloading transmit power, and the local computation frequency. The action space can be therefore expressed as $a_i = \{\rho_i, \hat{b}_i, p_i, f_i\}$. The actions of all UEs form a joint action, which can be mathematically denoted by ${\mathcal A} = \{a_i|i=1,\dots,I\}$.


At time step $t$, each UE is given a global reward $r_t$ based on the joint action ${\mathcal A}$. The reward is designed based on the objective function. As we aim to maximize the QoE of UEs, the value of reward should be positively correlated to the QoE. Meanwhile, we also set punishments for the executions that cannot meet the latency and accuracy constraints, which can be denoted by
\begin{equation}
r_i=
\begin{cases}
\rm{QoE}_i,&\text{Successful task execution},\\
t_{\rm{max}} - t_i, &\text{if (\ref{objective:c6}) not satisfied},\\
\epsilon_i - \epsilon_{\rm{min}},&\text{if (\ref{objective:c1}) not satisfied}, 
\end{cases}
\label{reward}
\end{equation}
and the sum reward $r_t = \sum^I_{i=1}r_i$.
When the tasks are completed or the time step ends, an extra reward is given to UEs for evaluating the execution performance of the whole task set. For the tasks finished before the max time step, a positive reward should be given for completing the tasks in advance. Otherwise, punishment should be given. Mathematically, the extra reward can be expressed by

\begin{equation}
r_{T}=
\begin{cases}
\xi_S(T-t_0),&t_0\leq T,\\
- \xi_F \sum \nolimits_{i\in I} d_i(T) E^{\mathcal Q}_0,&\rm{otherwise},
\end{cases}
\label{eq23}
\end{equation}
where $t_0$ is the task completed time step, $d_i(T)$ represents the length of the task queue $d_i$ at the max time step $T$, and $\xi_S$ and $\xi_F$ are constant values for the success extra reward and failure punishment. Hence, the sum reward for a completed trajectory $\tau$ can be expressed by $R(\tau) = \sum_{t=1}^{t_0} \gamma^{t-1} r_t + r_{T}$, where $\gamma \in (0,1)$ represents the discount factor, representing how much the future reward effect on the current state, $\tau$ is a sequence of states and actions which denotes the trajectory in the environment. 

Generally, all of the deep RL algorithms with continuous action space can be applied to (\textbf{P0}), e.g., deep deterministic policy gradient (DDPG), twin delay DDPG (TD3), asynchronous advantage actor-critic (A3C), PPO, etc. In this paper, we apply the PPO because of its robustness and good performance~\cite{PPO_2017}.

\subsection{MAPPO Training Process}

The local models are trained using the data in the local data set, which is called the replay memory in RL. At each training step $t$, the experience $e_{t, i} = (o_{t, i}, a_{t, i}, r_t, o_{t+1, i})$ acquired by $U_i$ is stored in $i$-th local replay memory ${\mathcal B}_i$. The objective function of the RL is to maximize the expected reward for each trajectory, which can be expressed by 
\begin{equation}
J\left (\pi_\theta\right) =
\mathbb{E}_{\tau\sim \pi_\theta(\tau)}\left [R(\tau)\right]= \int_{\tau} P(\tau|\pi_\theta) R(\tau),
\label{objective_function}
\end{equation}
where $\pi_\theta$ is the parameterized policy, $P(\tau|\pi_\theta) =P (o_0) \prod_{t=0}^{T-1} P(o_{t+1, i} | o_{t,i}, a_{t,i}) \pi_\theta(a_{t,i} | o_{t,i})$ represents the probability of the trajectory $\tau$, $P(o_{t+1, i} | o_{t,i}, a_{t,i})$ is the state transformation probability and $\pi_\theta(a_{t,i} | o_{t,i})$ is the action choice probability, and $P (o_0)$ is the probability of the initial state $o_0$. To optimize the policy, the policy gradient needs to be calculated, i.e., $\theta_{j+1} = \theta_j + \alpha \left. \nabla_{\theta} J(\pi_{\theta}) \right|_{\theta_j}$, where $\alpha$ is the learning rate or the learning step. The gradient of the policy can be expressed as
\begin{equation}
\nabla_{\theta}J(\pi_{\theta})\!=\!\mathbb{E}_{\tau \sim \pi_{\theta}(\tau)}\left[{\sum_{t=0}^{T} \nabla_{\theta} \log \pi_{\theta}(a_{t,i} |o_{t,i}) \Phi_{t,i}}\right],
\label{policy_gradient}
\end{equation}
where $\Phi_{t,i}$ is denoted as the value function while will be introduced later.

A feature of most on-policy policy gradient algorithms is that the trajectory $\tau$ is generalized by policy $\pi_\theta$ at all time steps. The gradient $\nabla_{\theta}J(\pi_{\theta})$ of the policy $\pi_\theta$ can only take the experience of current $\tau \sim \pi_{\theta}(\tau)$, while the previous experiences from other trajectories are discarded since they come from different policies. This characteristic disables the replay memory for training on-policy algorithms. Actually, for the proximal policies that differ gently from each other, previous experience should be considered to improve the training efficiency. To enable the PPO to learn from the proximal previous policy, the importance sampling technique should be adopted, which is given by Lemma~\ref{IS}.

\begin{lemma}[Importance sampling]
Given distributions $x\sim p(x)$ and $x\sim q(x)$, then $\mathbb{E}_{x\sim p}f(x) = \mathbb{E}_{x\sim q}\frac{p(x)}{q(x)}f(x)$.

\begin{proof}
Given $x\sim p(x)$, the expectation of $f(x)$ can be expressed as
\begin{align*}
       &\mathbb{E}_{x\sim p}f(x)\\
        &= \int_x p(x) f(x) \mathrm{d}x\\
        &= \int_x \frac{p(x)}{q(x)} q(x) f(x) \mathrm{d}x\\
        &= \mathbb{E}_{x\sim q}\frac{p(x)}{q(x)}f(x). \qedhere
\end{align*}
\label{IS}
\end{proof}
\end{lemma}
By invoking Lemma~\ref{IS}, the gradient $\nabla_{\theta}J(\pi_{\theta})$ of the policy $\pi_\theta$ could be calculated by the trajectory generated by $\pi_{\theta^{\prime}}$ as
\begin{equation}
\begin{split}
\nabla J_{\theta^{\prime}}^{\theta}
& = \mathbb{E}_{\tau \sim \pi_{\theta^{\prime}}(\tau)}\\
&\left[{\sum_{t=0}^{T} \frac{\pi_{\theta}(a_{t,i} |o_{t,i})}{\pi_{\theta^{\prime}}(a_{t,i} |o_{t,i})}
\nabla_{\theta} \log \pi_{\theta}(a_{t,i} |o_{t,i}) \Phi_{t,i}}\right].
\label{policy_gradient_old}
\end{split}
\end{equation}

To evaluate whether an action $a$ is good or not, the value function $\Phi_{t,i}$ can be denoted as 
\begin{equation}
    Q^{\pi_\theta}(o_{t,i},a_{t,i}) = \mathbb{E}_{\tau\sim \pi_\theta(\tau)}\left[R(\tau)| o_0 = o_{t,i}, a_0 = a_{t,i}\right],
\label{value_function}
\end{equation}
which is the expectation reward for taking action $a_{t,i}$ at state $o_{t,i}$. However, this function also relies on the state, which means the action-value function for an optimal policy under a bad state may be even lower than an arbitrary action under a better state, i.e., the state-value function $V^{\pi_\theta}(o_{t,i}) = \mathbb{E}_{\tau\sim \pi_\theta(\tau)}\left[R(\tau)| o_0 = o_{t,i}\right]$ need to be taken into consideration. Instead of comparing the action-value function directly, the advantage of an action $a$ compared with the other action under the state $o$ can be expressed by 
\begin{equation}
    A^{\pi_\theta}(o_{t,i},a_{t,i}) = Q^{\pi_\theta}(o_{t,i},a_{t,i}) - V^{\pi_\theta}(o_{t,i}).
\label{action_value_function}
\end{equation}
Additionally, if we need to consider the advantage function instead of the action-value function, the state influence should not affect the value of the policy gradient~\cite{GAE_paper}. The proof is given by Lemma~\ref{EGLP}.

\begin{lemma}[Expected Grad-Log-Prob Lemma]
Given $P^{\pi_\theta}$ is a parameterized probability distribution over a random variable $o$, then $\mathbb{E}_{o\sim P^{\pi_\theta}}\left[{\nabla_{\theta} \log P^{\pi_\theta}(o)}\right] = 0.$

\begin{proof} 
For all probability distributions, we have 
\begin{equation}
   \int_o P^{\pi_\theta}(o) = 1. 
\end{equation}
Therefore
\begin{align*}
       &\mathbb{E}_{o\sim P^{\pi_\theta}}\left[{\nabla_{\theta} \log P^{\pi_\theta}(o)}\right]\\
        &= \int_o P^{\pi_\theta}(o) \nabla_{\theta}\log P^{\pi_\theta}(o)\\
        &= \int_o \nabla_{\theta} P^{\pi_\theta}(o)\\
        &= \nabla_{\theta} \int_o P^{\pi_\theta}(o)\\
        &= 0.\qedhere
\end{align*}\end{proof}
\label{EGLP}
\end{lemma}
According to Lemma~\ref{EGLP}, we can derive that for the state-value function
$V^{\pi_\theta}(o_{t,i})$ that only related to the state, the gradient with the policy $\pi_\theta$ can be denoted as
\begin{equation}
    \mathbb{E}\left[{\nabla_{\theta} \log \pi_{\theta}(a_{t,i}|o_{t,i}) V^{\pi_\theta}(o_{t,i}))}\right]= 0.
\label{0_expectation}
\end{equation}
According to (\ref{value_function}), (\ref{action_value_function}), and (\ref{0_expectation}), the value function $\Phi_{t,i}$ can be denoted as the advantage function $A^{\pi_\theta}(o_{t,i},a_{t,i})$ with the state-value function $V^{\pi_\theta}(o_{t,i})$ elimination. Then (\ref{policy_gradient_old}) can be updated as
\begin{equation}
\begin{split}
\nabla J_{\theta^{\prime}}^{\theta}
& = \mathbb{E}_{\tau \sim \pi_{\theta^{\prime}}(\tau)}\\
&\left[{\sum_{t=0}^{T} {\rm{ratio}}_{t,i}
\nabla_{\theta} \log \pi_{\theta}(a_{t,i} |o_{t,i}) A^{\pi_\theta}(o_{t,i},a_{t,i})}\right],
\label{policy_gradient_ratio}
\end{split}
\end{equation}
where ${\rm{ratio}}_{t,i} =\frac{\pi_{\theta}(a_{t,i} |o_{t,i})}{\pi_{\theta^{\prime}}(a_{t,i} |o_{t,i})}$ represents the update step of the policy.

It is noted that the action-value function can be expressed by the temporal difference form as $Q^{\pi_\theta}(o_{t,i},a_{t,i}) = r_t + \gamma V^{\pi_\theta}(o_{t+1, i})$. However, the action-value function and the state-value function cannot be acquired directly from the experience $e_{t,i}$, and in deep RL approaches, we can use the NN to estimate the state-value function. In this way, we can express the estimated advantage function $\hat{A}^{\pi_\theta}(o_{t,i},a_{t,i}) = \delta^V_{t,i} = r_t + \gamma \hat{V}^{\pi_\theta}(o_{t+1,i}) -\hat{V}^{\pi_\theta}(o_{t,i})$. However, the bias for this estimation is high, which restricts the training and convergence performance. To overcome this issue, generalized advantage estimation (GAE)~\cite{GAE_paper} can be applied to estimate the advantage function for multi-steps and strike a tradeoff between the bias and variance. The GAE advantage function is denoted by
\begin{equation}
    A^{\rm{GAE}}(o_{t,i},a_{t,i}) =  \sum \limits _{l=0}^{T-t}(\lambda\gamma)^l\delta^V_{t+l,i},
\label{GAE_function}
\end{equation}
where $\lambda \in (0,1]$ is the discount factor for reducing the variance of the future advantage estimation.

The actor network is optimized by maximising  $L_{AC} = \mathbb{E}_{\tau \sim \pi_{\theta}(\tau)}\left[{\rm{ratio}}_{t,i} A^{\rm{GAE}}(o_{t,i},a_{t,i})\right]$. However, the large step could lead to an excessively large policy update, hence we can clip this step and restrict it. The clipped actor objective function is expressed by
\begin{equation}
    L^{\rm{Clip}}_t\!=\!\min\left(
{\rm{ratio}}_{t,i}\!\times\!A^{\rm{GAE}}(o_{t,i},a_{t,i}),\!
g(\delta,\!A^{\rm{GAE}}(o_{t,i},a_{t,i}))\right),
\label{clipped_loss}
\end{equation}
where 
\begin{equation}
g(\delta, A) =
\begin{cases}
(1 + \delta) A, &A \geq 0,\\
(1 - \delta) A & A < 0,
\end{cases}
\label{clip}
\end{equation}
in which $\delta$ is a constant value to control the policy update steps. The clip operation has been proved to improve the robustness of the model by openAI~\cite{PPO_2017}. 
The loss $L_{\rm{CR}}$ for the critic network is to minimize the gap between the estimated state-value function and the discount sum reward, which can be expressed by
\begin{equation}
    L^{\rm{CR}}_t =\left\Vert \hat{V}^{\pi_\theta}(o_{t,i}) - R_t \right\Vert^ 2,
\label{TD}
\end{equation}
where $R_t = \sum^T_{l=t} \gamma^{l-t} r_l  + r_{T}$ represents the discount future reward from time step $t$, which can be estimated by $\hat{R}_t = r_t + \gamma\hat{V}^{\pi_\theta}(o_{t+1,i}).$

Combining the objective of the actor network and critic network, we can express the overall objective as
\begin{equation}
L=\arg \max_{\theta} \mathbb{E}_{t}\left[L^{\rm{Clip}}_t-c^{\rm{CR}}L^{\rm{CR}}_t+c^{E}E\right],
\label{overall_loss}
\end{equation}
where $E$ represents an entropy bonus to ensure sufficient exploration, $c_1$ and $c_2$ are weight parameters for the estimation of value function and entropy, respectively.


\section{Numerical Results}
\label{sec:numerical_results}

\begin{table}[t]
\begin{center}
\caption{Simulation Parameters}
\label{tab1}
\begin{tabular}{|c|c|}
\hline
\textbf{Parameter}&
\textbf{Value} \\ 
\hline
Number of UEs $I$&
4\\
\hline
Carrier frequency&
6 $\rm{GHz}$\\
\hline
Sum bandwidth $BW$&
5 $\rm{MHz}$\\
\hline
noise power spectrum density&
-160 $\rm{dBm/Hz}$\\
\hline
Transmit power range&
$(18,24) \rm{dBm}$\\
\hline
Static and circuit power&
$20 \rm{dBm}$\\
\hline
Sum bandwidth $BW$&
5 $\rm{MHz}$\\
\hline
Text semantic encoder output dimension&
16\\
\hline
Image semantic encoder output dimension&
32\\
\hline
Number of processors of the local GPU&
1024\\
\hline
Local GPU frequency range&
$(0.64, 1.28) \rm{GHz}$\\
\hline
Number of processors of the remote GPU&
13824\\
\hline
Remote GPU frequency&
$1.41 \rm{GHz}$\\
\hline
Minimum accuracy requirements $\epsilon_{min}$&
0.9\\
\hline
Execution latency constraint  $t_{\rm {max}}$&
$1 \rm{s}$\\
\hline
Training episode of the MAPPO algorithm&
1000\\
\hline
Testing episode of the MAPPO algorithm&
100\\
\hline
Batch step of the MAPPO&
256\\
\hline
Learning rate&
2e-7\\
\hline
{Advantage discount factor $\lambda$}&
0.98\\
\hline
Reward discount rate $\gamma$&
0.95\\
\hline
{Maximum policy update step $\delta$}&
0.3\\
\hline
Critic network loss weight $c^{\rm{CR}}$&
0.5\\
\hline
Policy entropy bonus weight $c^E$&
0.005\\
\hline
\end{tabular}
\end{center}
\end{table}

\begin{figure}[t]
\subfigure[Fixed QoE preference.]{
\begin{minipage}[t]{\linewidth}
\includegraphics[width=0.99\columnwidth]{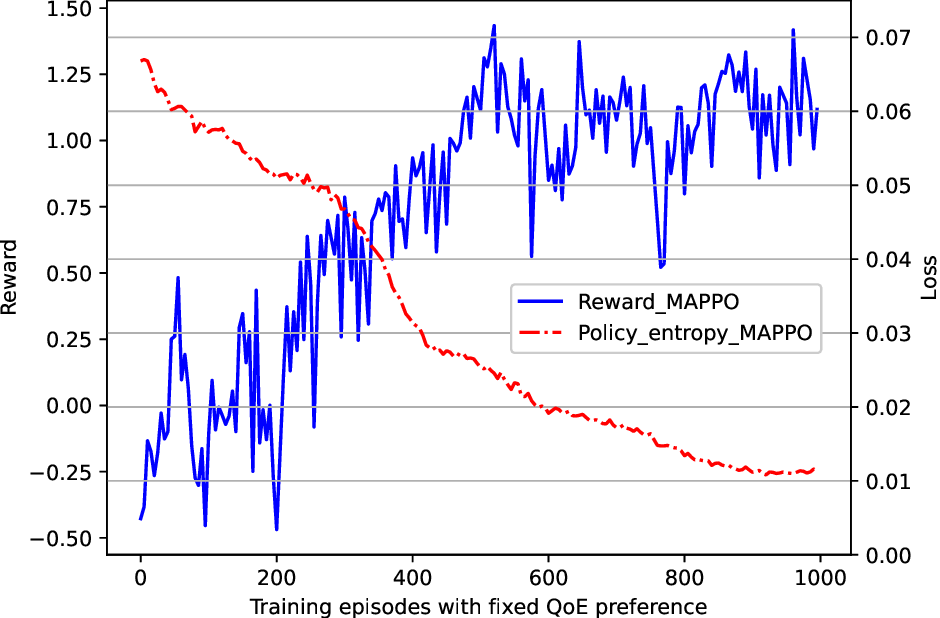}
\end{minipage}%
}%

\subfigure[Dynamic QoE preference.]{
\centering
\begin{minipage}[t]{\linewidth}
\includegraphics[width=0.99\columnwidth]{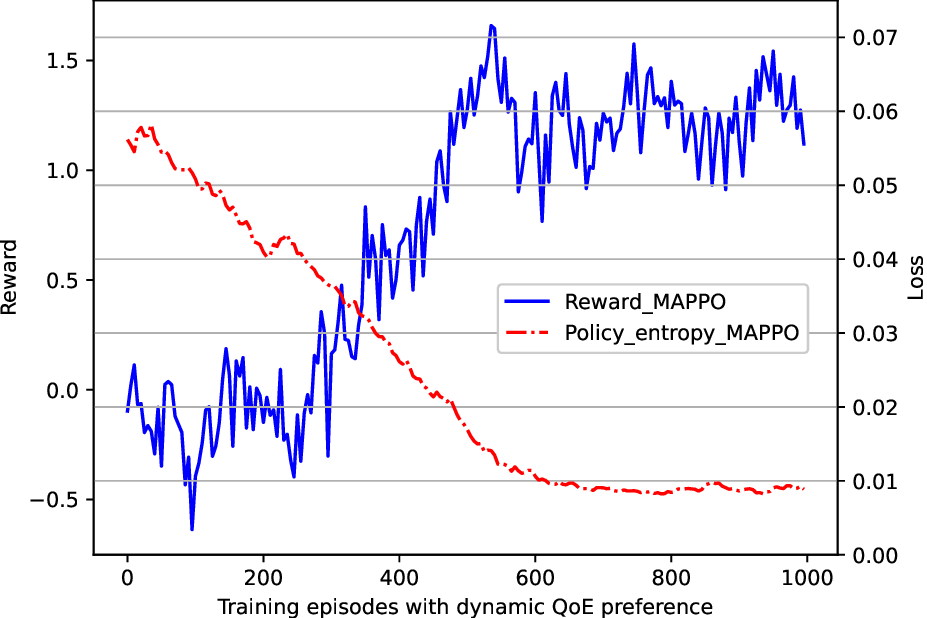}
\end{minipage}%
}%
\caption{Training performance comparison of the proposed algorithm and benchmarks in three different scenarios.}
\label{training_performance}
\end{figure}

\begin{figure*}[t]
\subfigure[Machine translation tasks.]{
\begin{minipage}[t]{0.33\linewidth}
\includegraphics[width=0.99\columnwidth]{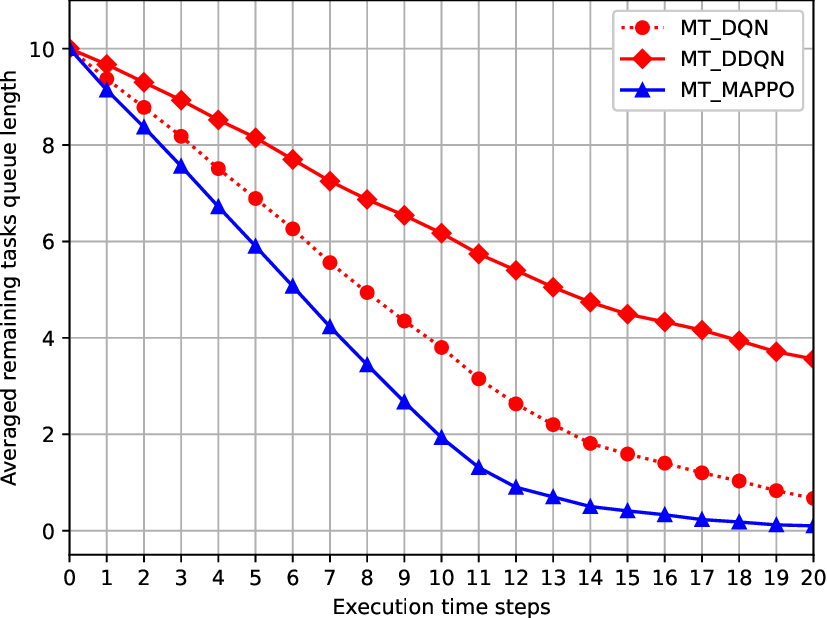}
\end{minipage}%
}%
\subfigure[Image retrieval tasks.]{
\centering
\begin{minipage}[t]{0.33\linewidth}
\includegraphics[width=0.99\columnwidth]{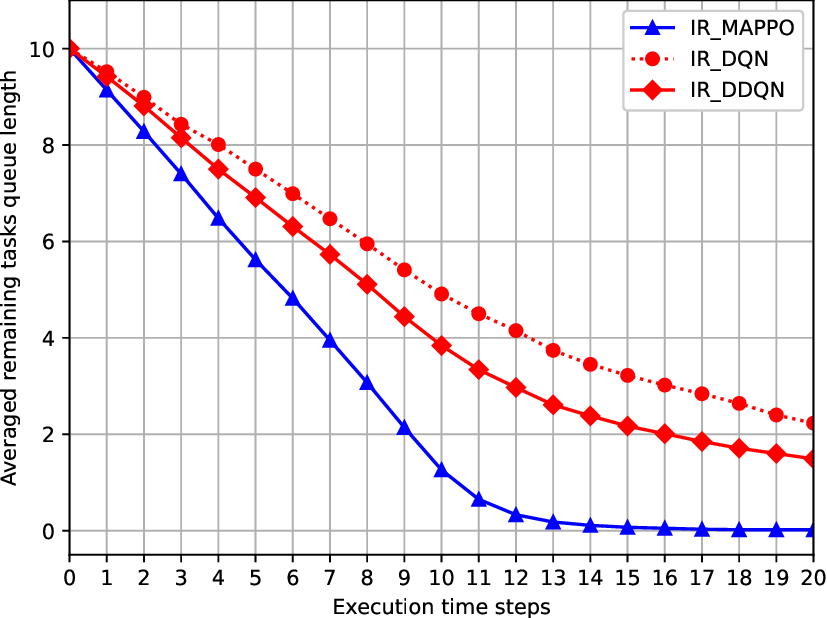}
\end{minipage}%
}%
\subfigure[VQA tasks.]{
\centering
\begin{minipage}[t]{0.33\linewidth}
\includegraphics[width=0.99\columnwidth]{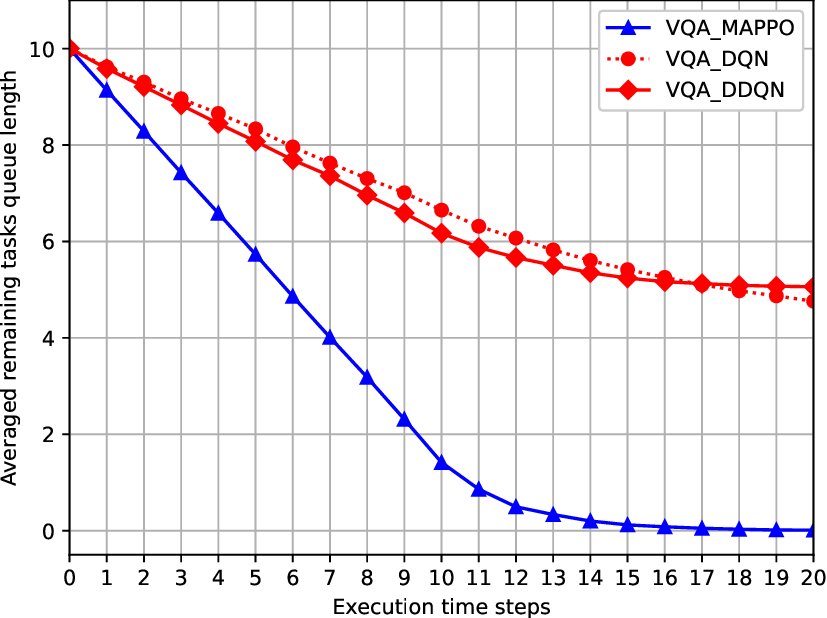}
\end{minipage}%
}%
\caption{Average queue length of the multi-modal tasks with fixed QoE preference. The results are averaged by 100 task sets with random user distributions.}
\label{Queue_length}
\end{figure*}

\begin{figure*}[t]
\subfigure[Machine translation tasks.]{
\begin{minipage}[t]{0.33\linewidth}
\includegraphics[width=0.99\columnwidth]{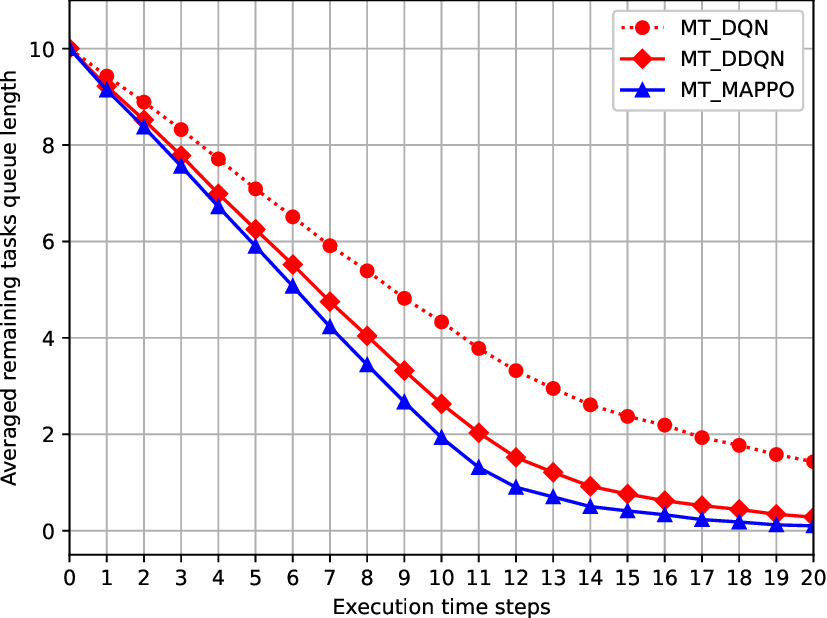}
\end{minipage}%
}%
\subfigure[Image retrieval tasks.]{
\centering
\begin{minipage}[t]{0.33\linewidth}
\includegraphics[width=0.99\columnwidth]{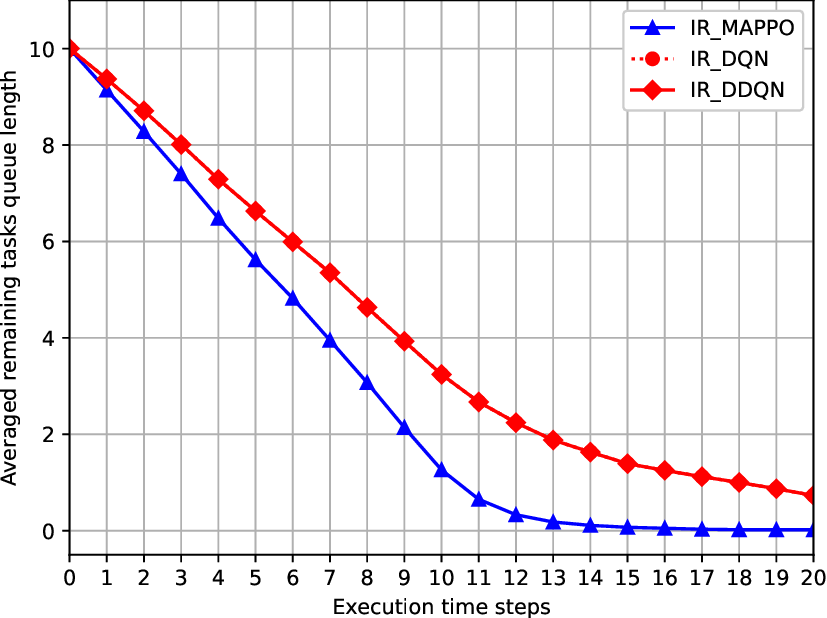}
\end{minipage}%
}%
\subfigure[VQA tasks.]{
\centering
\begin{minipage}[t]{0.33\linewidth}
\includegraphics[width=0.99\columnwidth]{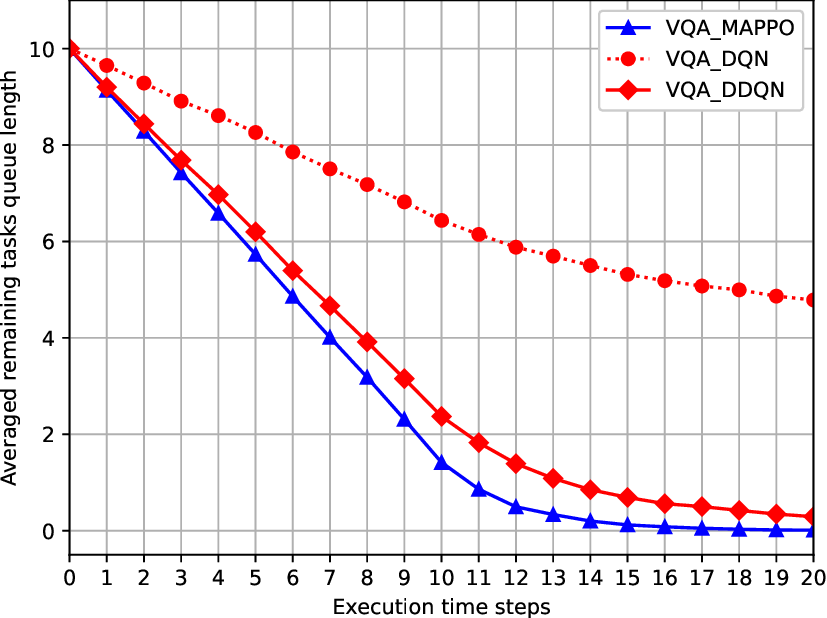}
\end{minipage}%
}%
\caption{Average queue length of the multi-modal tasks with dynamic QoE preference. The results are averaged by 100 task sets with random user distributions.}
\label{Dynamic_queue_length}
\end{figure*}



\subsection{Implement Details}

The text semantic encoders for the machine translation tasks and VQA tasks are based on the public implementation of the Transformer~\cite{vaswani2017attention} model
with $F^{\mathcal T} = 6$ Transformer encoder layers and decoder layers. The image encoders for the image retrieval tasks and VQA tasks are based on the public implementation of the DeiT-small model~\cite{touvron2021training}, where contains $F^{\mathcal I} = 6$ Transformer encoder layers and decoder layers. Each task set for machine translation tasks includes $d^{\mathcal T} = 10$ paragraphs, and each paragraph contains 10 sentences with average length $S^{\mathcal T} = 20$. For each image retrieval task, $d^{\mathcal I} = 10$ images are added to the image task queue and processed one by one. Each image is decomposed into $16 \times 16$ patches and an extra $<$CLS$>$ token, i.e., $S^{\mathcal I} = 197$. 

Based on the above design of encoders and decoders, we implement the semantic communication system based on the trained DeepSC models~\cite{9830752} and the accuracy mappings, i.e., $ f^{\mathcal T}$, $f^{\mathcal I}$, and $f^{\mathcal B}$, can be obtained by testing the pre-trained DeepSC models under different SNRs. Total $I = 4$ UEs are deployed within the service range of the ES, including one with machine translation tasks, one with image retrieval tasks, and two with VQA tasks.

The pathloss model is set according to~\cite{3gpp_38.901}. Moreover, the simulation settings for the wireless channels, the specification of GPUs for the ES and local UEs, and the parameters for training the MAPPO models are listed in TABLE \uppercase\expandafter{\romannumeral1}.

\subsubsection{Training Settings and Convergence Performance}
This section demonstrates the training performance of the proposed MAPPO algorithm for joint resource optimization. First, we need to determine how to design a practical QoE preference scheme. The QoE preference on latency and energy consumption may vary among UEs. We simulate two different QoE schemes to cope with the practical requirements of UEs. In the first scheme, we consider the QoE preference to be fixed and identical among UEs, where $\beta^f = 0.5$ represents that UEs have the same preference on the latency requirement and energy consumption requirement. However, the preference of UEs may differ over time. For instance, a UE with a fully charged battery may care more about the latency, while a user with a lower battery life may need lower energy consumption. Hence, the second scheme considers the current battery life of UEs and we defined the QoE preference for energy consumption as 
\begin{equation}
    \beta^d = 1 - \frac{B}{2B_{\rm{max}}},
\label{QoE_preference}
\end{equation}
which represents the UEs have the same preference for energy consumption and latency when the battery is fully charged, while caring more about the energy consumption when the battery life is lower. 

The convergence of the proposed MAPPO has been demonstrated in Fig. \ref{training_performance}. It is clear that the proposed algorithm can achieve convergence at about 600 episodes, and keeps at near-optimal performance over the rest of the training stages. It is noted that for the dynamic QoE preference scheme, the reward is less stable than the fixed one. The reason is that the models are trying to learn a dynamic policy with the dynamic QoE scheme. Unlike the fixed policy where the UEs can refer to the previous trajectory with the same preference and objective function, the dynamic scheme is hard to learn from the previous experience, leading to slower convergence. Nevertheless, the proposed scheme can still achieve good convergence, which can be verified by the policy entropy indicator. The policy entropy is a metric to measure the randomness of a policy for policy gradient based RL algorithms. The decreasing trends of fixed and dynamic schemes verify the convergence performance of the proposed algorithm.

\subsubsection{Benchmarks and Task Execution Tests}
To verify the performance of the proposed MAPPO algorithm, we compare it with the following benchmarks. 
\begin{itemize}
\item {\textbf{Exhaustive search}: The upper bound of the proposed problem (\textbf{P0}) is hard to acquire. To evaluate the near-optimal results, we apply the exhaustive search of the discrete power and frequency levels to find the optimal discrete solutions.}
\item {\textbf{Deep Q-network (DQN)}: This algorithm is based on the traditional DQN and multi-agent RL algorithm. The transmit power and computation frequency are converted to discrete levels, as the DQN can only deal with discrete action space.}
\item {\textbf{Double deep Q-network (DDQN)}: An improved version of DQN, which addresses the overestimation issue of the DQN. DDQN only deals with discrete action space.}
\item {\textbf{Local}: All tasks are performed at local UEs.}
\item {\textbf{Remote}: All tasks are offloaded to the ES and performed by the ES.}
\item {\textbf{Random}: The offloading policy, the transmit power, the channel requirement, and computation frequency are randomly chosen by each UE.}
\end{itemize}

Figs.~\ref{Queue_length} and~\ref{Dynamic_queue_length} present the remaining task length for the UEs with machine translation tasks, image retrieval tasks, and VQA tasks, respectively. We compare the proposed MAPPO algorithm and the DQN and DDQN benchmarks. Each simulation result denotes the averaging queue length of 100 testing episodes with 100 random task sets and random user distributions, while the curve for the VQA tasks represents the averaging queue length of the two UEs with VQA tasks. It is clear that the proposed MAPPO algorithm can execute all of the tasks within the time step limits, while the DQN and DDQN algorithm behaves worse, especially for fixed QoE cases. Meanwhile, the proposed MAPPO algorithm achieves less averaging queue length and behaves stably. In most cases, the task sets can be processed within 15 time steps, which provides a stable execution performance guarantee. However, the conventional RL algorithms, i.e., DQN and DDQN algorithms cannot process the whole task queue within the time step limitation, providing a bad guarantee of the accomplishment of the task sets. Unlike the DQN or DDQN with discrete action space and optimizes the value function itself, the proposed MAPPO algorithm optimizes the policy as well as the value function simultaneously, enabling it a higher possibility to find a better continuous solution.

\subsection{QoE Optimization Performance}
Previous simulations demonstrate that the MAPPO outperforms the conventional deep RL algorithms with both the fixed QoE and dynamic QoE preference schemes. In the following simulations, we only apply the scheme with a more practical setting with dynamic QoE preference, which puts forward a higher test for the robustness of the algorithm. For the semantic task offloading scheme, the average number of semantic symbols, i.e., the output dimension of the semantic encoders per input unit is $k^{\mathcal Q}$, which is important since it determines the computational complexity as well as the communication costs. A lower output dimension represents a higher compression rate of the source information, which reduces the transmission overhead, but may affect the decoding performance and decrease the task execution accuracy. Nevertheless, $k^{\mathcal Q}$ cannot be optimized as a variable since the semantic communication models should be pre-trained and freeze during the resource optimization process. Hence, here we only investigate the system QoE preference over different output dimensions, to determine the most suitable $k^{\mathcal T}$ and $k^{\mathcal I}$ for the proposed semantic task offloading system. 

\begin{figure}[t]
\centering
\includegraphics[width=0.99\columnwidth]{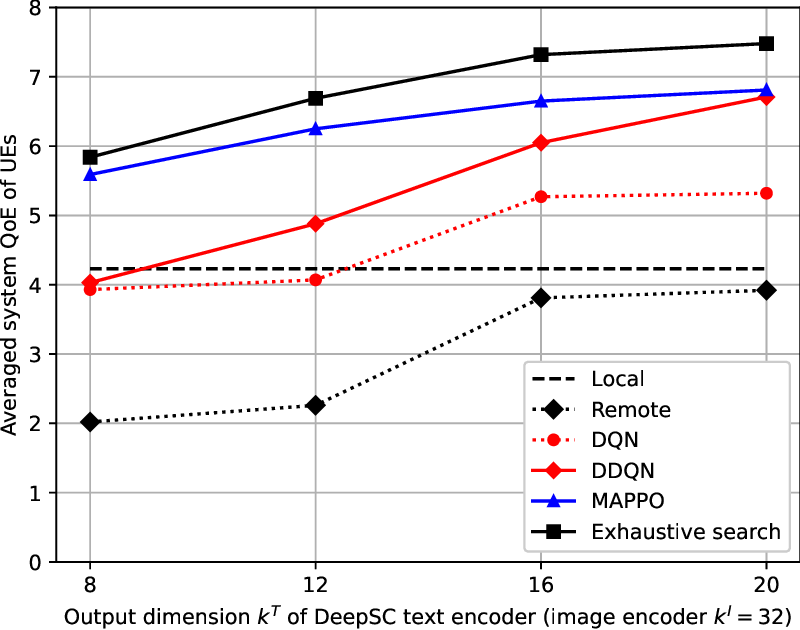}
\caption{QoE with the output dimension of semantic text encoders $k^{\mathcal T}$.}
\label{averaged_QoE_with_text}
\end{figure}

\begin{figure}[t]
\centering
\includegraphics[width=0.99\columnwidth]{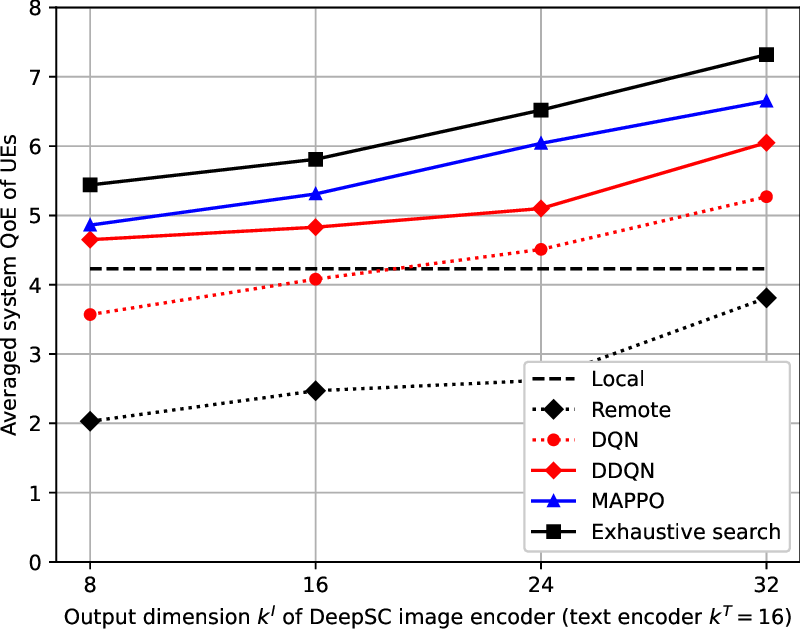}
\caption{QoE with the output dimension of image semantic encoders $k^{\mathcal I}$.}
\label{averaged_QoE_with_image}
\end{figure}

Fig.~\ref{averaged_QoE_with_text} and~\ref{averaged_QoE_with_image} demonstrate the system sum QoE over different dimensions of the text semantic encoders and image encoders. It is noted that the local benchmark remains stable at 4.23, which represents that the output dimension does not influence the QoE of the local execution scheme. The reason lies in that the computing complexity of the output layer is much smaller than the complexity of the Transformer layers, hence the output dimension does not affect the overall FLOPs of the encoders $F_{\rm{encoder}}$. Meanwhile, the task execution accuracy is always 1 for local cases due to (\ref{cost}) and ({\ref{qoe}}). The sum system QoE takes the task completion time step into account and remains stable regardless of different text semantic encoder output dimensions. 

In Fig.~\ref{averaged_QoE_with_text}, we can see that the system sum QoE increases with the number of output dimension of text semantic encoder $k^{\mathcal T}$. This is because the semantic information of the tasks can be decoded successfully, and the accuracy of the task execution improved with a larger $k^{\mathcal T}$. Meanwhile, the QoE remains stable when output dimension $k^{\mathcal T}$ increases from 16 to 20, this is because the UEs require more communication resources to offload the extracted semantic features and the execution cost increases. While for the image semantic encoders as shown in Fig.~\ref{averaged_QoE_with_image}, the QoE improves with the larger output dimension $k^{\mathcal I}$, representing that the task execution accuracy gain dominates the QoE performance, hence we choose $k^{\mathcal T} = 16$ and $k^{\mathcal I} = 32$ as TABLE~\ref{tab1}.

An issue occurs with the decreasing of the output dimensions $k^{\mathcal T}$ and $k^{\mathcal I}$. The lower output dimension leads to worse decoding performance, thus reducing the task execution accuracy and the system QoE. When the task execution accuracy cannot satisfy the constraint~\ref{objective:c1}, the task will fail and a punishment reward will be given to UEs. If a UE can hardly receive a positive reward, which is called the sparse reward issue, it cannot learn from a good policy. Hence, when $k^{\mathcal T}$ is relatively small, i.e., equal to 8 and 12, or $k^{\mathcal I}$ equal to 8 and 16, the performance gaps between the reinforcement learning algorithms and the upper bound found by the exhaustive search method are large. Even though, the proposed MAPPO algorithm outperforms the DQN, DDQN, and other benchmarks, achieving near-optimal QoE.

\begin{figure}[t]
\centering
\includegraphics[width=0.99\columnwidth]{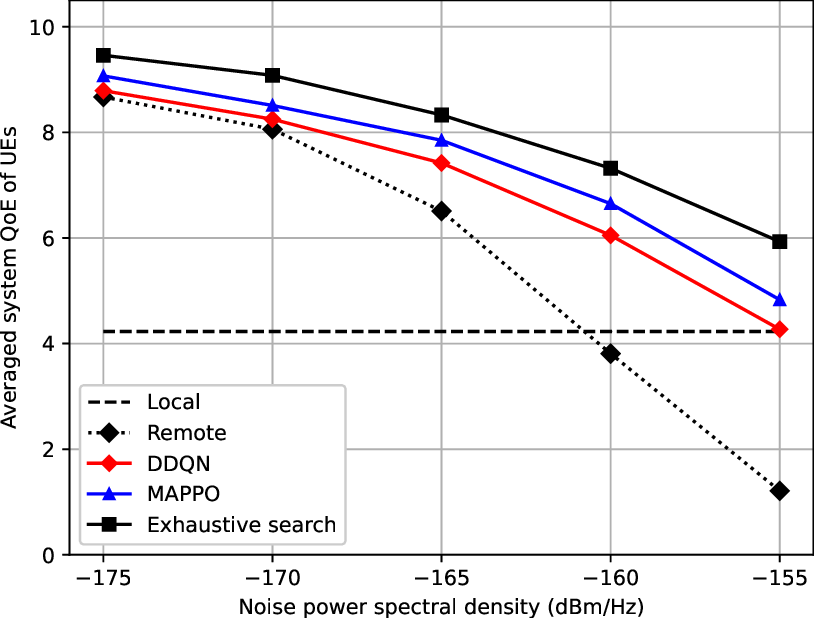}
\caption{QoE with noise power spectrum density.}
\label{averaged_QoE_with_noise}
\end{figure}

In Fig.~\ref{averaged_QoE_with_noise}, we further investigate the QoE performance over different noise strengths, i.e., the task offloading QoE under different SNRs. The noise power spectrum density ranges from $-175$ $\rm{dBm/Hz}$ to $-155$ $\rm{dBm/Hz}$. It is observed that the performance of the proposed MAPPO algorithm always outperforms the DDQN algorithm, local execution, and remote execution schemes. With the noise power increasing, the accuracy of the task executed remotely decreases significantly, and the system QoE also drops with the decreasing accuracy. When the noise power spectrum density is equal to $-155$ $\rm{dBm/Hz}$, the DDQN algorithm cannot be well-trained due to the sparse reward issue, with even worth QoE performance compared to the local execution scheme. Meanwhile, the QoE performance for all schemes will consider offloading the tasks to the ES when the noise power spectrum density is low. When the noise power spectrum density is equal to $-175$ $\rm{dBm/Hz}$, the gap is narrowed between the upper bound found by the exhaustive search algorithm and the fixed remote execution scheme, and the curves for MAPPO and DQN algorithms also achieve similar performance with the remote scheme.

\section{Conclusion}
\label{sec:conclusion}
In this paper, a semantic-aware multi-modal task offloading system has been investigated and the quality of experience of users has been maximized by the proposed proximal policy optimization based multi-agent reinforcement learning algorithm. The semantic information of the machine translation tasks, the image retrieval tasks, and visual question answering are extracted by the text semantic encoders and image semantic encoders, which can be offloaded to the edge servers or decoded and executed locally. Numerical analysis verifies the necessity of task offloading, and the simulation results demonstrate the proposed algorithm outperforms the benchmarks as it finishes the task sets with fewer time steps. The proposed algorithm also achieves a higher quality of experience under different settings of the semantic encoders and the noise power, further verifying its robustness.

\appendices
\section{Quantity Analysis for Semantic Offloading System}
The computational complexity of semantic encoder and decoder (by Floating-point operations, FLOPs) is derived in this section. We use the text semantic encoder as the example and assume that the input of the machine translation tasks is a sentence $\boldsymbol{x^{\mathcal T}}=\{x[0], x[1], \dots, x[S^{\mathcal I}]\}$, where $S^{\mathcal T}$ is the number of words in a sentence. Here, we use $S = S^{\mathcal T}$ to simplify the expressions, and the computational complexity for the image semantic encoder can be derived similarly. For the Transformer layers, $d_k$, $d_v$, $d_{\rm{model}}$, $L$, $h$, and $B$ represent the size of the query vector, the value vector, the word embedding dimension, the number of the Transformer layers, the number of attention head, and the batch size, respectively. Hence, the input size of the DeepSC model should be $(B,S,d_{\rm{model}})$. The computation of the encoder includes the following stages. 

Firstly, we need to calculate the query, key, and value matrices by multiplying the embedding and matrices, i.e., $Q = X\cdot W^Q$, $K = X\cdot W^K$, and $V=X\cdot W^V$. According to~\cite{vaswani2017attention}, we set $H = d_k=d_v=d_{\rm{model}}/h$, thus the dimension of these matrices is $(d_{\rm{model}}, H)$, hence, the FLOPs of stage 1 can be expressed by
\begin{equation}
F_1 = 3 \cdot  2 \cdot  B\cdot S \cdot  d_{\rm{model}} \cdot  H =
6B\cdot S\cdot h\cdot H^2.
\end{equation}
Secondly, the query matrices are multiplied by key matrices, i.e., $Q\cdot K^T$, where the FLOPs can be expressed by $F_{2,1} = 2B\cdot H\cdot S^2$. Then the results are softmax and multiplied by Value matrices, where the FLOPs can be denoted by $F_{2,2} = 2B\cdot H\cdot S^2$, then the FLOPs in stage 2 should be
\begin{equation}
F_2 = 4B\cdot h\cdot H\cdot S^2.
\end{equation}
Note that the results from multi-heads need to be summed by multiplying a weight matrix $W^0$, thus the overall FLOPs for a single attention layer should be
\begin{equation}
\begin{aligned}
   F_{\rm{atten}} &= h\cdot (F_1+F_2)+2B\cdot S\cdot h^2\cdot H^2 \\
   &= 8B\cdot S\cdot h^2\cdot H^2+4B\cdot S^2\cdot h^2\cdot H.
\end{aligned}
\end{equation}
At the output layer of each Transformer layer, the attention vector needs to pass a feedforward layer, where the FLOPs can be denoted by $F_{\rm{FFN}} = 2B\cdot S\cdot h\cdot H^2$. Here is noted that the computational complexity for the attention layers is much more than the FFN layer.

For the semantic encoder of DeepSC, the semantic information needs to be passed by a dense layer for channel encoding. Suppose that the number output of the dense layer is $n_{\rm{out}}$, and the FLOPs for the dense layer is $F_{\rm{dense}} = 2B\cdot S\cdot h\cdot H\cdot n_{\rm{out}}$. Thus, the total number of FLOPs for the DeepSC encoder is denoted as
\begin{equation}
F_{\rm{encoder}} = L(F_{\rm{atten}} + F_{\rm{FFN}}) + F_{\rm{dense}}.
\end{equation}

The FLOPs for the decoder of DeepSC can be expressed similarly. However, there are two attention layers in each Transformer layer for the decoder, one for self-attention and another for encoder-decoder attention. Additionally, the output layer of the decoder is a classification layer to determine which words need to choose from the vocabulary set, thus the FLOPs of the output layer should be $F_{\rm{vocab}} = 2B\cdot S\cdot H\cdot n_{\rm{vocab}}$, where $n_{\rm{vocab}}$ represents the size of the vocabulary set.

Hence, the total number of FLOPs for the decoder of the DeepSC can be denoted as
\begin{equation}
F_{\rm{decoder}} = L(2F_{\rm{atten}} + F_{\rm{FFN}}) + F_{\rm{vocab}}.
\end{equation}

As the analysis above, the computational complexity for the attention layers is much more than other layers, i.e., $F_{\rm{atten}} \gg F_{\rm{FFN}}$. The decoder requires twice as many attention layers as the encoder, which inspires us to offload the computational cost of the semantic decoder to the edge servers.

\bibliography{Reference}
\clearpage

\end{document}